\documentclass{aa}
\usepackage{graphicx}
\usepackage{txfonts}
\usepackage{graphicx}
\usepackage[normalem]{ulem}
\usepackage{color}

\begin{document}

   \title{Mapping the outer bulge with RRab stars from the VVV Survey}
   \titlerunning{Mapping the outer bulge with RRab stars from the VVV Survey}
   \authorrunning{Gran et al.}

   %\subtitle{}

\author{F.~Gran\inst{1}\fnmsep\inst{2},
        D.~Minniti\inst{3}\fnmsep\inst{2}\fnmsep\inst{4},
        R.~K.~Saito\inst{5},
        M.~Zoccali \inst{1}\fnmsep\inst{2},
        O.~A.~Gonzalez \inst{6,7},
        C.~Navarrete\inst{1}\fnmsep\inst{2},
        M.~Catelan\inst{1}\fnmsep\inst{2},\\
        R.~Contreras Ramos\inst{2}\fnmsep\inst{1},
        F.~Elorrieta\inst{8}\fnmsep\inst{2},
        S.~Eyheramendy\inst{8}\fnmsep\inst{2},
        A.~Jord\'an\inst{1}\fnmsep\inst{2}}

\institute{Instituto de Astrof\'isica, Pontificia Universidad Cat\'olica
           de Chile, Av. Vicu\~na Mackenna 4860, 782-0436 Macul, Santiago,
           Chile\\
           \email{fgran@astro.puc.cl}
      \and
     The Millennium Institute of Astrophysics (MAS), Santiago, Chile
      \and
     Departamento de Ciencias Fisicas, Universidad Andres Bello,
          Republica 220, Santiago, Chile
            \and
     Vatican Observatory, V-00120 Vatican City State, Italy
            \and
           Universidade Federal de Sergipe, Departamento de F\'isica,
       Av. Marechal Rondon s/n, 49100-000, S\~ao Crist\'ov\~ao, SE, Brazil
     \and
     European Southern Observatory, Alonso de Cordova 3107, Vitacura,
         Santiago, Chile
     \and
     Institute  for  Astronomy,   University  of  Edinburgh,  Royal
     Observatory, Blackford Hill, Edinburgh, EH9 3HJ, UK
     \and
     Departmento de Estad\'istica, Facultad de Matem\'aticas, Pontificia
     Universidad Cat\'olica de Chile, Av.\ Vicu\~na Mackenna 4860,
     7820436 Macul, Santiago, Chile}

\date{Received Sept XX, 2015; accepted Oct XX, 2015}

  \abstract
{  The  VISTA Variables  in  the V\'ia  L\'actea  (VVV)  is a  near-IR
  time-domain survey of the Galactic bulge and southern plane.  One of
  the main goals  of this survey is to reveal the  3D structure of the
  Milky Way  through their variable stars.  Particularly  the RR Lyrae
  stars have  been massively  discovered in the  inner regions  of the
  bulge ($-8^\circ  \lesssim b \lesssim -1^\circ$)  by optical surveys
  such as OGLE and MACHO but leaving an unexplored window of more than
  $\sim  47$ sq deg ($-10.0^\circ  \lesssim \ell \lesssim +10.7^\circ$
  and $-10.3^\circ  \lesssim b  \lesssim -8.0^\circ$) observed by the
  VVV Survey. }
   { Our goal is to characterize  the RR Lyrae stars in the {\it outer
       bulge}   in  terms  of   their  periods,   amplitudes,  Fourier
     coefficients,  and  distances,  in   order  to  evaluate  the  3D
     structure of  the bulge in this area.   The distance distribution
     of RR Lyrae stars will be  compared to the one of red clump stars
     that is known to trace a X-shaped structure in order to determine
     if these two different stellar  populations share the same
     Galactic distribution.}
  % methods  heading (mandatory)  
{ A search for RR Lyrae stars  was performed in more than $\sim 47$ sq
  deg  at low  Galactic  latitudes ($-10.3^\circ  \lesssim b  \lesssim
  -8.0^\circ$).  In  the procedure the $\chi^2$ value  and analysis of
  variance  (AoV)  statistic methods  were  used  for determining  the
  variability and periodic features of the light curves, respectively.
  To prevent  misclassifications, the  analysis was performed  only on
  the fundamental mode RR Lyrae stars (RRab) due to similarities found
  in the near-IR  light curve shapes of contact  eclipsing binaries (W
  UMa) and  first overtone RR Lyrae  stars (RRc).  On  the other hand,
  the  red  clump stars  of  the  same  analyzed tiles  were  selected
  applying  cuts in  the color-magnitude  diagram and  restricting the
  maximum distance to  $\sim 20$ kpc to construct  a similar catalogue
  in  terms of distances  and covered  area compared  to the  RR Lyrae
  stars.}
{ We report the detection of  more than 1000 RR Lyrae ab-type stars in
  the VVV Survey located in  the outskirts of the Galactic bulge. Some
  of  them   are  possibly  associated  with   the  Sagittarius  Dwarf
  Spheroidal Galaxy.  We calculated colors, reddening, extinction, and
  distances of the  detected RR Lyrae stars in  order to determine the
  outer  bulge 3D  structure.  Our  main result  is that,  at  the low
  galactic latitudes mapped here, the RR Lyrae stars trace a centrally
  concentrated   spheroidal  distribution.    This  is   a  noticeably
  different spatial distribution to the  one traced by red clump stars
  known  to  follow  a  bar  and  X-shape  structure.We  estimate  the
  completeness   of   our   RRab   sample  in   $80\%$   for   $K_{\rm
    s}\lesssim15$~mag.}
   {}


   \maketitle

\section{Introduction}
\label{sec:intro}

Big  astronomical  surveys are  changing  the  way  we understand  the
formation, structure and evolution of  our Galaxy. Among those, only a
few  have been  able to  access  the inner  regions of  the Milky  Way
because  of  the effects  of  severe  crowding  and high  interstellar
extinction of these dense  Galactic regions.  Near- and mid-IR surveys
such as 2MASS, GLIMPSE,  and UKIDSS-GPS \citep{2mass, glimpse, ukidss}
have helped to overcome  the extinction problem covering the innermost
regions  of the Galaxy,  but the  lack of  multiple-epoch observations
within  those  surveys  prevents  us  from using  them  to  study  and
characterize   the   large  number   of   variable   sources  in   the
bulge.  Optical time-domain  surveys  such as  OGLE,  MACHO, and  EROS
\citep{ogleiv,  macho, eros}  have partially  solved this  problem but
unfortunately   the   high   extinction   found  towards   the   bulge
line-of-sight  restricts  them   from  accurately  map  the  innermost
regions.

In response  to these  limitations, the VISTA  Variables in  the V\'ia
L\'actea (VVV) ESO public survey \citep{minniti10} provides a near-IR,
multi-epoch photometric coverage of the inner Galaxy ($\rm -10^{\circ}
\lesssim  \ell  \lesssim  10^{\circ}$,  $\rm  -10^{\circ}  \lesssim  b
\lesssim 5^{\circ}$).   The large near-IR coverage of  the VVV survey,
high spatial resolution, and depth  of the survey enables the capacity
to perform  studies globally across the entire  inner Galaxy, reaching
larger distances  than it  has ever been  possible before.   The first
stage of the VVV Survey provided full-coverage, multi-color photometry
of the inner 520 sq. deg. of  the Galaxy.  These data was used for the
construction      of     2-D      and     3-D      extinction     maps
\citep{gonzalez11,gonzalez13, schultheis14},  and metallicity gradient
maps \citep{gonzalez13} of the Galactic bulge.

One  of the main  scientific goals  of the  VVV Survey  is to  build a
global  3D map  of the  Milky Way  using well  known  primary distance
indicators. In this  context, the first epoch of  VVV observations has
been used  to investigate  the shape of  the bulge using  the observed
magnitude  of red-clump  giant  stars as  distance indicators.   Bulge
studies using  red-clump stars have helped unveiling  the global shape
of the  stellar bar, confirming that  the Milky Way hosts  a peanut or
X-shape bulge \citep{wegg13, saito12b}.

On the other hand, the  ongoing variability campaign of the VVV survey
now  allows us  to investigate  the shape  of the  inner  Galaxy using
variable  stars as  distance  estimators. Variable  star searches  are
expected   to  yield  many   more  candidates   in  the   near  future
\citep{catelan13a, catelan13b}, allowing us to measure the extinctions
and distances  along the line of  sight, providing another  3D view of
the  inner Milky  Way \citep{dekany13,dekany15}.   RR Lyrae  stars are
particularly interesting  in this context  as they allow us  to trace,
unequivocally,   the   oldest   stellar   component  of   the   Galaxy
\citep{dekany13, catelan15}.  Interestingly, the distance distribution
of RR Lyrae  stars found by \cite{dekany13} follows  a different shape
than the one traced by  red-clump stars.  While the distances obtained
from red-clump stars  trace tightly the position angle  of the bar, as
well as  the distance split  along the minor  axis due to the  far and
near arms  of the X-shaped  bulge, distances and radial  velocities to
the  RR  Lyrae population  from  \cite{dekany13} and  \cite{kunder16},
respectiely, appear to follow a spheroidal distribution instead of the
stellar bar traced by RC stars.

In this present study we perform the search of RR Lyrae
stars  using VVV  data continuing  the analysis  which was  started by
\cite{gran15},  extending  the  work   to  28  more  VVV  tiles  ({\it
  b201}-{\it b228}).  These regions have  been not been covered by the
OGLE  survey   yet,  therefore  the   RR  Lyrae  presented   here  are
particularly important  in this context. This is where  the X-shape
bulge  becomes  most  prominent   making  it  the  ideal  location  to
investigate  how different  are  the structures  traced  by these  two
populations. We  calculate their  distances and compare  their spatial
distribution with respect to those derived from red-clump stars.

\section{Observations}
\label{sec:obs}

The VVV  Survey is  a public  ESO near-IR survey  that is  mapping the
inner Milky Way,  including the inner halo, the  bulge and an adjacent
section  of  the disk  with  the VISTA  4m  telescope  at ESO  Paranal
Observatory \citep{minniti10}. The survey covers an area of 562 sq deg
in total, and the VVV database now contains $ZYJHK_{\rm s}$ photometry
of  about one  billion sources  on the  VISTA system  for  which 2MASS
coordinates have been  used to construct the coordinate  system, and a
variability   campaign  in   the  $K_{\rm   s}$-band  \citep{saito12a,
  hempel14}. See \cite{gran15} for  more details of the instrument and
their spatial configuration on the Galactic bulge and disk.

In this  analysis we used  data covering more than  $\sim
  47$ sq  deg in the outer bulge  ($-10.0^\circ \lesssim
  \ell  \lesssim  +10.7^\circ$ and  $-10.3^\circ  \lesssim b  \lesssim
  -8.0^\circ$).  This  area corresponds to  the VVV tiles  {\it b201}
through {\it b228},  obtained between April 2010 and  August 2014 with
60-62 epochs  in all the  selected tiles.  We use  aperture photometry
applied  to  the  stacked  images  (e.g. ``tiles"),  provided  by  the
Cambridge             Astronomical             Survey             Unit
(CASU)\footnote{\url{http://casu.ast.cam.ac.uk/vistasp/}}  and setting
the  minimum number  of epochs  per star  analyzed to  30 in  order to
achieve  a better  frequency  analysis  and avoid  gaps  in the  light
curves.

\subsection{Detection and classification of RR Lyrae stars}
\label{sec:class}

We selected  variable candidates by  analyzing the $\chi^2$  value for
all  the available  time series,  considering the  mean error-weighted
magnitude as  the model  (e.g.  a non-variable  star will  have values
close to 0). A similar analysis was presented in \cite{carpenter01} to
detect variable candidates.  If  this value exceeds the imposed cutoff
of $\chi^2  = 2$ \citep[see][]{gran15},  the time-series periodicities
are tested by the analysis  of variance (AoV) statistic \citep{aov} in
the  RR Lyrae  stars period  range ($0.2  \leq P  \text{  (days)} \leq
1.2$).    After  this   process   the  light   curves  were   visually
classified.

\begin{figure}
\centering
\includegraphics[scale=0.5]{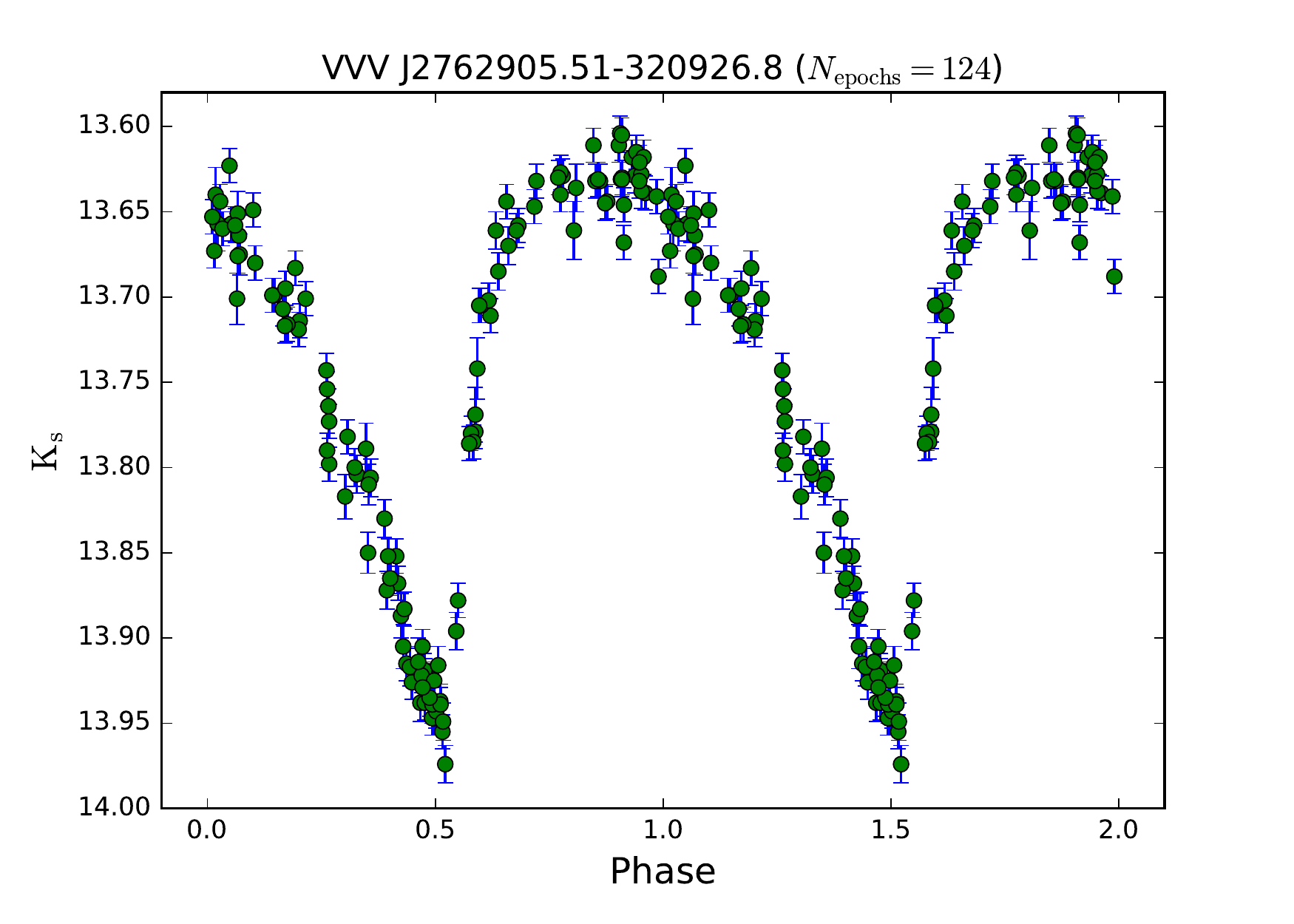}
\caption{RR  Lyrae star  in the  overlap of  two adjacent  tiles ({\it
    b208} and {\it  b222}). The light curve has  the maximum number of
  epochs in our sample ($62 \times 2 = 124$).
}
\label{fig:overlap}
\end{figure}

We  repeated the  classification process  over the  28  analyzed tiles
({\it  b201}-{\it b228})  and check  if there  were duplicates  in our
catalogues.   RR  Lyrae  stars  in  the intersection  areas  are  also
important  in   order  to  check  the  parameters   derived  from  two
independent light-curves.  The tiling  pattern produces about $7\%$ of
overlapping  areas  between  the  tiles  \citep{saito12b},  thus  took
advantage of the duplicated RR  Lyrae stars in the overlapping regions
by combining their data. Fig.~\ref{fig:overlap} shows a RRab star with
the  maximum number of  epochs found  in the  intersection of  the VVV
tiles {\it b208}  and {\it b222}.  For the  overlapping RR Lyrae light
curves,  the derived  periods,  amplitudes, and  mean magnitudes  were
compared, resulting in a distribution  close to 0 within the errors in
the parameters.

In  this process we  assign a  label to  the RRab  stars due  to their
narrow  period range  ($\sim 0.4  \leq  P \text{  (days)} \leq  1.2$),
near-IR  amplitude  ($0.2 \lesssim  A_{\rm{K_{\rm  s}}} \text{  (mag)}
\lesssim 0.5$) and characteristic  asymmetric light curve shape (e.g.,
Figure~\ref{fig:overlap}).  As  reported by \cite{alonso-garcia15}, in
near-IR bands the quantity  of features to classify different variable
types are fewer than in the optical regime.  Therefore, because of the
light curves of  RRc stars in the near-IR mimic  the behavior of other
variable classes such as W UMa contact binaries and long-period SX Phe
pulsating  variables, likely  RRc  stars ($P  \text{ (days)}  \lesssim
0.4$) are  not under analysis here.   In addition to  the human expert
classification  described above,  we run  the light  curves  through a
machine-learned    classifier   specifically    developed    for   the
classification of RRab in the VVV Survey. The classifier is based on a
set of  features extracted from  each light curve following  a similar
approach  to that  of \cite{debosscher07}  and  \cite{richards11}, and
will be described  elsewhere (Elorrieta et al 2016,  in prep). We will
use a similar classifier in the  near future to produce a catalogue of
VVV variable sources  classified using automated procedures \citep[for
  more details see][]{catelan13b, angeloni14}.

One of the 28 tiles explored  is obliterated by the presence of a very
bright star, resulting  in fewer RR Lyrae discovered.  Tile {\it b205}
contains the star  $\eta$ Sgr (HD 167600) which is  very bright in the
near-IR with $K_{\rm  s} \sim -1.55$ mag. Such a  bright star not only
saturates the  detector, but also  causes reflections that  affect the
flat fields,  and the resulting  mosaic of this tile  contains regions
that are not suitable for variability searches. This is the reason why
tile {\it b205} contains fewer  RRab stars ($N_{\rm{RRab}} = 31$) than
the rest of the tiles ($N_{\rm{RRab}} \sim 37$ on average).

 Our  RRab  light  curves  have  $60-62$  data-points  with  a  median
 magnitude of $K_{\rm s}$=14.2~mag  ($12.1 \lesssim K_{\rm s} \lesssim
 16.3$).  At this  magnitude level the completeness of  the VVV source
 catalogues is high, with about $95\%$ of detection efficiency in less
 crowded    fields    such    as    the   outermost    bulge    region
 \citep{saito12a}. On the other  hand, experiments of signal detection
 rates based on  VVV data for RRab stars  reach about $90\%$ detection
 when  applied on  light curves  with  $60$~epochs \citep{catelan13b}.
 Therefore, we  can estimate  the completeness of  our RRab  sample as
 good as $80\%$ for $K_{\rm s}\lesssim15$~mag, with no expected trends
 along both axes, since crowding and extinction are similar across the
 analyzed area.  At fainter magnitudes the completeness is smaller and
 making it  difficult to find the  most distant RR  Lyrae, for example
 the  ones that  may  belong to  the  Sgr dwarf  galaxy.  However,  we
 identify a few Sgr RR Lyrae candidates (see Section~\ref{sec:dist}).

We have  also checked the  completeness of our catalogue  by comparing
our findings  with the RR Lyrae found  by OGLE in a  small fraction of
our area which overlaps an  OGLE IV field \citep{ogleIVRRL}. There are
22 RR Lyrae stars with $-10.3^\circ \lesssim b \lesssim -8.0^\circ$ in
the OGLE IV catalogue of which we will only focus on the 13 RRab stars
present.  In  our catalogue there  are 8 matches  within $d <  1''$ in
tiles {\it  b220} and  {\it b221}.  Three  of the five  remaining RRab
stars were not analyzed by our algorithm due to non-stellar photometry
flags  or less  epochs than  the minimum  required and  there  were no
matches  at $1''$  for the  last two  RRab stars  in the  area  in our
catalogue.  With this corrections our completeness respect to the OGLE
survey is at  least 80\%.  Certainly not all of the  RRab stars in the
catalogue are new discoveries. We match our catalogue with the General
Catalogue of Variable Stars \citep[GCVS;][]{GCVS}, with a total of 207
matches.  VVV IDs  and the respective GCVS names  for matching objects
are presented in  Appendix A.  We emphasize the fact  that none of our
classified RRab  stars have tagged eclipsing  binaries counterparts in
the GCVS,  despite that we do  not discard minor  contamination due to
eclipsing binaries that can mimic  RRab stars. Finally, 27 of the RRab
stars in tile b201 have already been reported by \cite{gran15}.

\section{Results}

\begin{figure}
\centering
\includegraphics[scale=0.4]{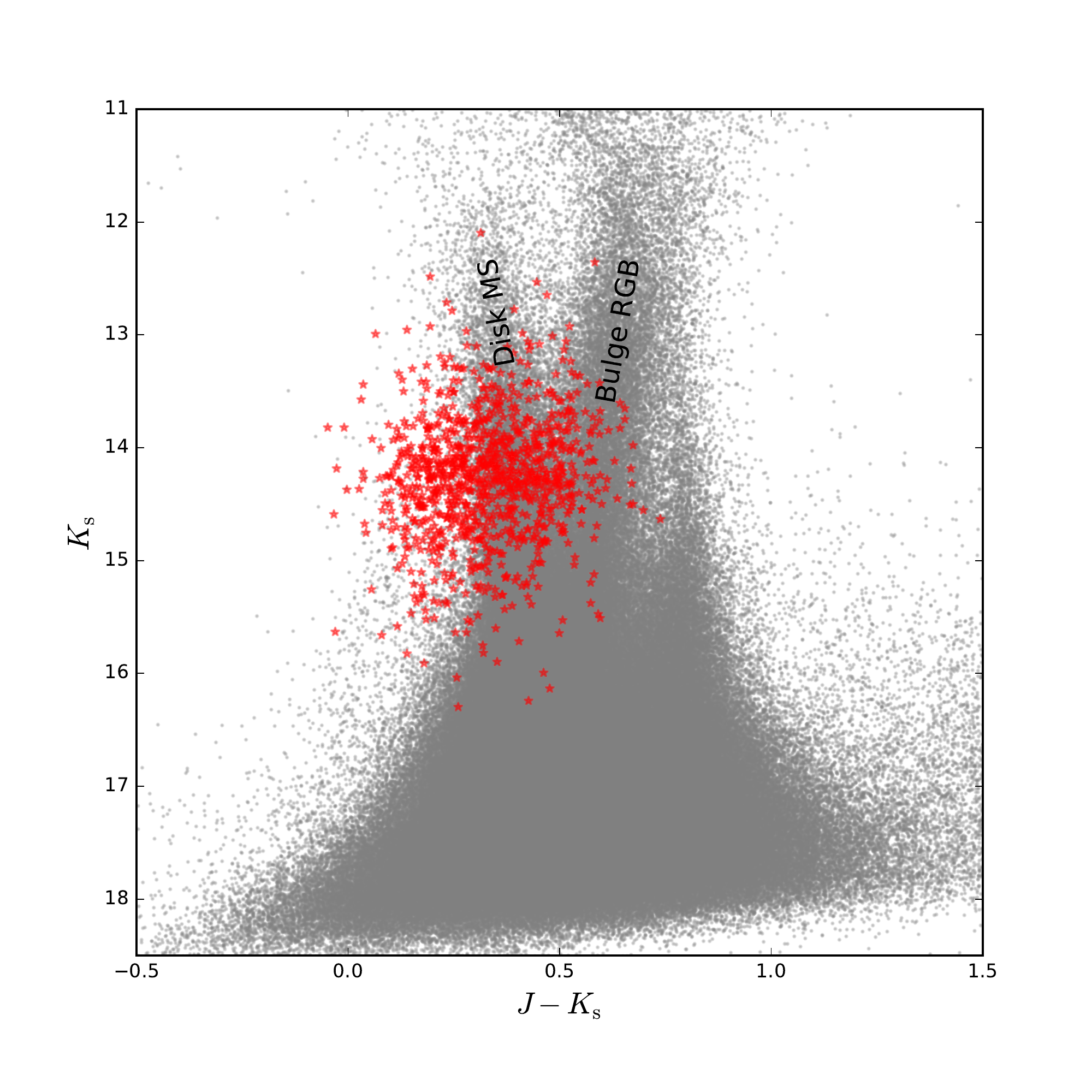}
\caption{ $K_{\rm s}$  {\it vs.}  ($J-K_{\rm s})$ CMD  of the complete
  catalogue of RR Lyrae stars  (red circles) compared with the sources
  in  tile {\it  b201} as  background.   The CMD  shows two  prominent
  features, the disk main sequence (MS) and the bulge red giant branch
  (RGB), that are identified in the figure.}
\label{fig:cmd}
\end{figure}

After accounting for the duplicates,  a total of 1019 RRab remained in
our sample. The final catalogue is presented in the Appendix A. In the
first step  we characterized  this sample in  terms of  its calculated
magnitude-weighted $\langle K_{\rm{s}}  \rangle$, $\langle J \rangle -
\langle  K_{\rm{s}} \rangle$  color, periods,  amplitudes, light-curve
shapes, and,  coordinates.  Fig.~\ref{fig:cmd}  shows the $J  - K_{\rm
  s}$  color-magnitude  diagram  (CMD)   for  the  complete  RR  Lyrae
catalogue with  tile {\it  b201} as a  comparison field.   Circles and
stars represent the overlapping and single detection stars of the RRab
sample,  respectively.  The  RR Lyrae  stars lie  in a  wide  range of
mean-$K_{\rm s}$ magnitudes due  to their distance distribution in the
Galaxy, but the $J-K_{\rm s}$  color is limited between $\sim 0.0$ and
$0.6$, similar values to those reported by \cite{gran15}.

\begin{figure}
\centering
\includegraphics[scale=0.38]{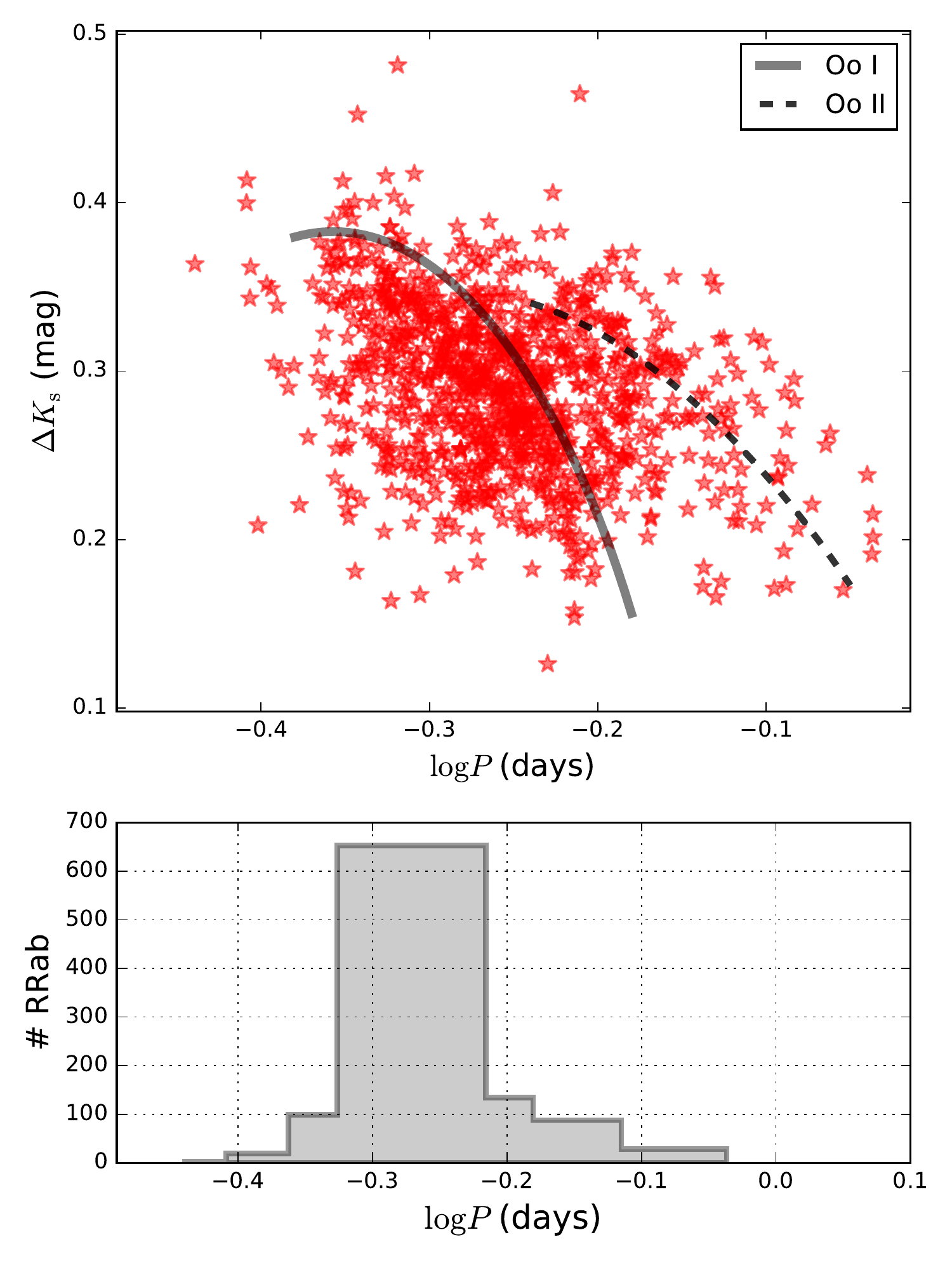}
\caption{ Top  panel: Bailey diagram  of the complete  RRab catalogue.
  The   OoI    (solid)   and   OoII   (dashed)    lines   derived   by
  \cite{navarrete15} are shown.  Bottom panel: period histogram of the
  1019 RRab  stars with bins  adapted by the Bayesian  Block algorithm
  \citep{scargle14}   through   the   {\tt   astroML}   implementation
  \citep{astroML}.}
\label{fig:bailey}
\end{figure}

Besides the locus on the CMD,  the RR Lyrae stars can be identified by
their position  on the Bailey diagram  \citep{bailey02}, which relates
the  period  and amplitude  of  the RR  Lyrae  stars,  and the  period
distribution  of the  entire sample  (Fig.~\ref{fig:bailey}).  In this
diagram  we can  derive  that  our RR  Lyrae  stars are  predominantly
Oosterhoff Type I (OoI) with a minor composition of Oosterhoff Type II
(OoII). We derive this composition with the Oosterhoof reference lines
traced by \cite{navarrete15}.

\begin{figure}
\centering
\includegraphics[scale=0.4]{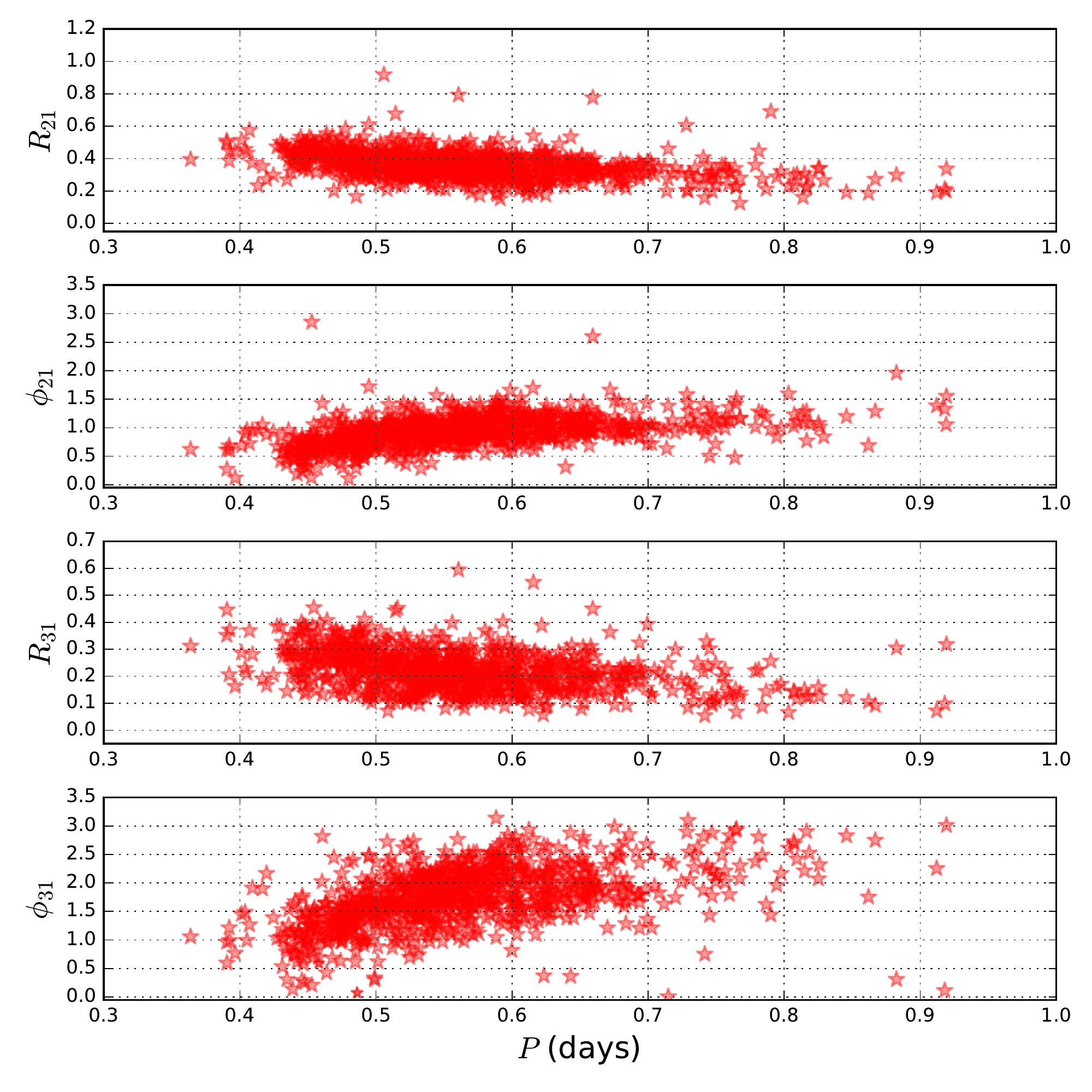}
\caption{ Top to  bottom: $R_{21}$, $\phi_{21}$, $R_{31}$
  and $\phi_{31}$ coefficients of  a Fourier series (sine based) using
  the DFF routine.}
\label{fig:fourier}
\end{figure}

In  addition  to  the  period and  amplitude,  another  characteristic
feature  of the  RRab  stars is  the  light-curve shape,  that can  be
described by a Fourier series.  A sine decomposition up to sixth order
was performed with  the Direct Fourier Fitting (DFF)  routine given by
\cite{tff}.  Fig.~\ref{fig:fourier} shows  the  $R_{21}$, $\phi_{21}$,
$R_{31}$ and  $\phi_{31}$ coefficients  as function of  the
period. All the  Fourier components tend to be  clustered in a limited
region  in this  space \cite[for  reference see  Fig.  6 of][]{deb10}.
There were some outliers in the distributions (e.g.: RRab with $R_{21}
> 0.6$ or $\phi_{21} > 2$) which were visually inspected, finding some
gaps in the light-curve that impact in the final value.

\begin{figure*}
\centering
\includegraphics[scale=0.5]{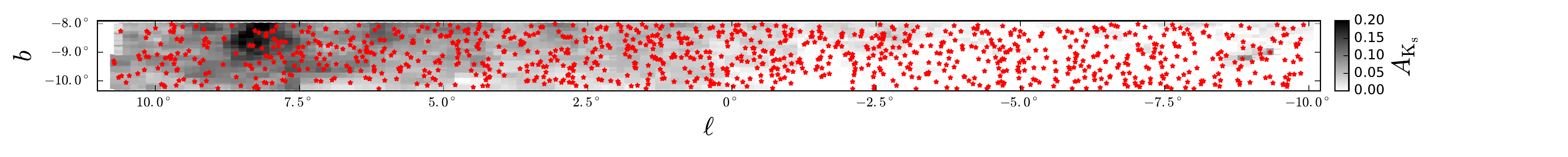}
\caption{Spatial distribution in Galactic coordinates ($\ell$, $b$) of
  the RRab stars found in this work.}
\label{fig:lb}
\end{figure*}

The spatial  distribution in Galactic coordinates of  the catalogue is
shown in  Fig.~\ref{fig:lb}. The  observations span only  2$^\circ$ in
$b$  but  more  than  20$^\circ$  in $\ell$,  resulting  in  the  very
elongated figure  shape.  Although there  are no globular  clusters in
the  analyzed  area  according  to  the  \cite{GGC}  catalogue,  their
presence in  nearby regions  could bias the  number of RR  Lyrae stars
found. This possible  effect in our catalogue was  investigated on the
three closest globular clusters to  our sample of RRab stars. NGC 6656
is the only  cluster that have associated RR  Lyrae stars according to
the               \citet[2015              edition              online
  catalogue\footnote{\url{http://www.astro.utoronto.ca/~cclement/read.html}}]{clement},
but  the closest  variable is  $10'$  further from  the cluster  tidal
radius  ($r_t  \approx  30'$)  given   by  the  2010  version  of  the
\cite{harris} catalogue.  NGC 6624  and 6637 are considered metal-rich
clusters  with  [Fe/H] values  of  $-0.63$  and $-0.77$,  respectively
\citep{6624,  6637}.   Both clusters  develop  a  very red  horizontal
branch, which is the reason why  they are not known to have associated
RR Lyrae stars.

\subsection{Distances and the 3D view of the outer bulge}
\label{sec:dist}

One  of the main  goals of  the VVV  Survey is  to trace  the Galactic
structure using variable  stars in order to make  the most complete 3D
view  of the  central  regions of  our  Galaxy \citep{minniti10}.  The
primary distance  indicators are  the RR Lyrae  stars due to  the high
number  density present in  the bulge  area \citep{ogleIVRRL}  and the
tight period--luminosity  (P--L) relation that they  follow in near-IR
bands \citep{longmore90,  catelan04}.  To obtain  the distance values,
in first place  we must calculate the reddening  and extinction values
to  the individual  variables.  The former  quantity  can be  obtained
through  the   difference  between  the   mean-apparent  and  absolute
magnitudes of our RRab stars, given by,

\begin{equation}
  E(J-K_{\rm s}) = (J-K_{\rm s}) - (J-K_{\rm s})_0 = (J-K_{\rm s}) - (M_J-M_{K_{\rm s}}),
\end{equation}
where $(J-K_{\rm  s})_0$ is the intrinsic  color of our  RRab star and
$M_X$ the absolute magnitude in the $X$-band. In our analysis we adopt
the P--L  relations derived  by \cite{alonso-garcia15} to  recover the
absolute  magnitudes of the  RR Lyrae  stars in  the $J$-  and $K_{\rm
  s}$-bands with  $\log Z = [\text{Fe/H}]  - 1.765$, based  on a solar
metallicity of $Z_\odot =  0.017$ \citep{catelan04}.  To calculate the
$J$-band mean magnitudes for the stars in our catalogue we performed a
linear  regression   between  the  $J$-   and  $K_{\rm{s}}$-band  mean
magnitudes of the RRab stars of $\omega$ Centauri studied by Navarrete
et al.   (2016, submitted).  This  analysis is needed because  the VVV
Survey  only  provides  one  observation  in  the  $ZYJH$-bands.   The
resulting fit  is given  by $\langle J  \rangle = 0.93  \times \langle
K_{\rm s} \rangle + 1.26$.  As expected, the residuals are centered in
$0$ with  a dispersion of  $0.03$ mag.  This  allows us to  derive the
reddening on a star-by-star  basis, and additionally the extinction of
each     RRab    star,     by    adopting     an     extinction    law
\citep[e.g.,][]{cardelli89}.

\begin{figure}
\centering
\includegraphics[scale=0.45]{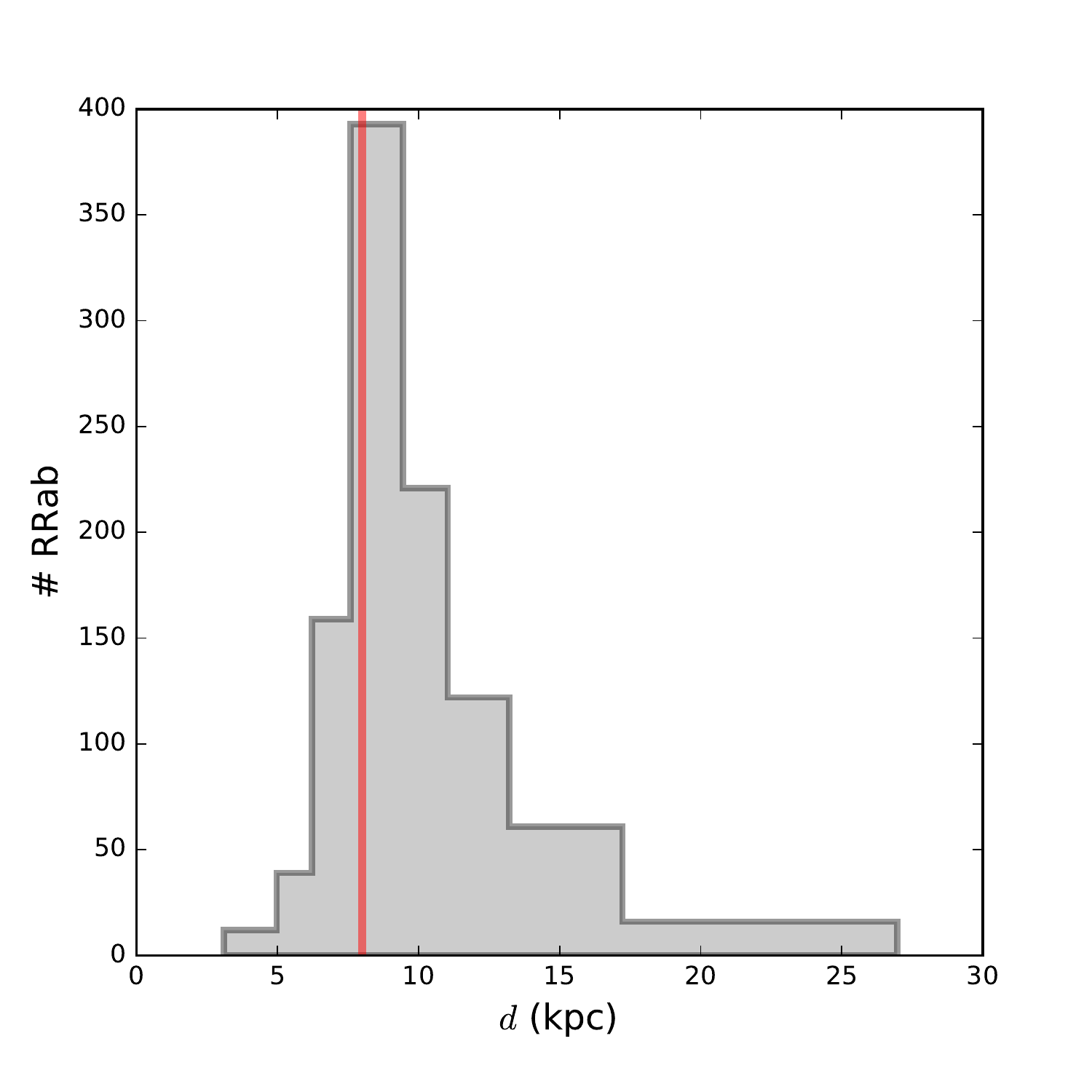}
\caption{Distribution of distances of the RR Lyrae found. The vertical
  line represents the Galactic  center derived by \cite{dekany13} with
  OGLE-III RR Lyrae stars of $R_0 \approx 8.33$ kpc.}
\label{fig:dist}
\end{figure}

At this point we calculate the distances given by,

\begin{equation}
  \log{d} = 1 + 0.2(K_{\rm{s,0}} - M_{K_{\rm s}}),
\end{equation}
with   $d$    the   individual   distance   in   pc    to   our   RRab
stars. Fig.~\ref{fig:dist} shows the  distribution of distances of the
RRab stars  in our  catalogue.  The vertical  line corresponds  to the
Galactic center  distance derived in  \cite{dekany13} with a  value of
$R_0 \approx 8.33$ kpc. Our  distances have a maximum frequency around
$R_0$ where the center of the distribution is, and an asymmetric shape
towards  the far  side of  the bulge  because the  volume  observed is
bigger due to  the cone effect. According to  their distances, some of
the RR Lyrae stars may belong to the Sagittarius dwarf spheroidal (Sgr
dSph) galaxy (e.g., distances around $20$ kpc).  \cite{kunder09} place
the core of  the Sgr dSph galaxy $\sim  22 - 27$ kpc from  the Sun but
$\sim 4^\circ$ away from our analyzed region. Even taking into account
that Sgr RR  Lyrae stars are mixed with the  Milky Way halo variables,
some  RR  Lyrae  stars  found  towards  these  coordinates  have  been
associated to the dwarf  galaxy by MACHO \citep{alard96, alcock97} and
OGLE \citep{ogleIVRRL}.

\begin{figure*}
\centering
\includegraphics[scale=0.7]{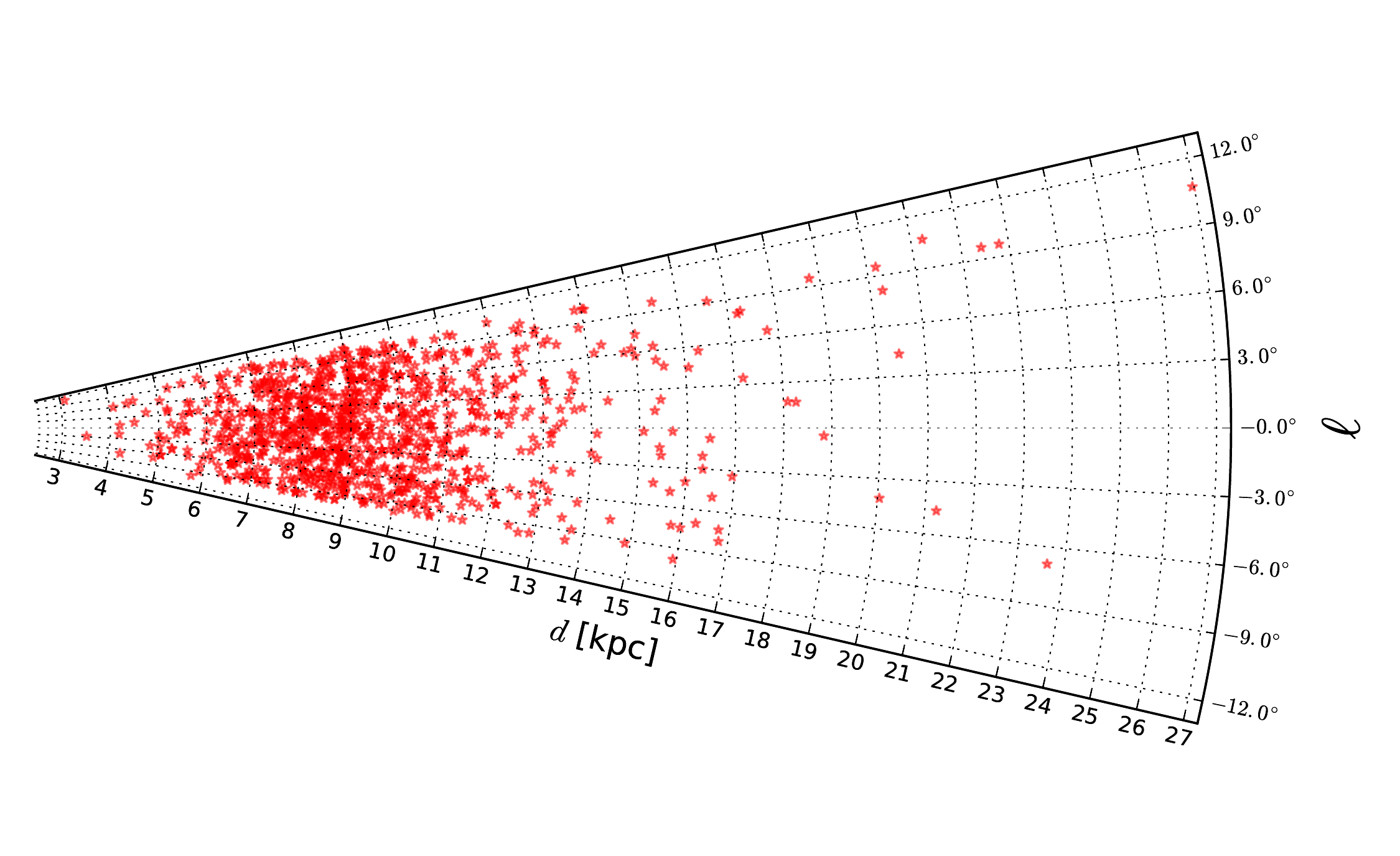}
\caption{Cone-view  ($d,\ell$) of  the analyzed  area in  the Galactic
  bulge.  The sample  is  concentrated around  the  projection of  the
  Galactic center.}
\label{fig:cone}
\end{figure*}

The elongated shape  of the analyzed area allow  us to approximate the
observation volume by a circular sector, projecting the
$b$  coordinate.   Fig.~\ref{fig:cone}  shows distances  and  Galactic
longitude  in this  line-of-sight  circular  projection.
The RR  Lyrae stars  tend to stay  near the projected  Galactic center
distance ($d \approx 8$ kpc)  and the previously mentioned Sgr dSph RR
Lyrae candidates are  clearly visible in the $16  \leq d \text{ (kpc)}
\leq 22$ and $\ell \geq 6^\circ$ zone.

\subsection{Trace or not to trace: the X-shape problem}

Many efforts have been made to study the 3D structure of the Milky Way
through its  stellar content. One important distant  indicator are the
pulsating variable stars (e.g.,  RR Lyrae and Cepheids, among others),
but besides  this method, the red  clump (RC) stars were  also used in
near-IR single epoch studies to derive accurate distances to the Milky
Way  edge  \citep{minniti11},   bulge  \citep{alves00}  or  the  Large
Magellanic  Cloud \citep{alves02}. This  feature of  the RC  stars has
been used recently to discover  the X-shape structure of the Milky Way
    \citep{mcwilliam10,    nataf10,   saito11,
    wegg13}   that   contains   a    bar   in   its   central   parts
\citep{rattenbury07, gonzalez11}.  This structure probably
  vanishes  with  decreasing  metallicity  of  stars, and  it  is  not
  expected in an  old stellar population \citep{ness12}.
It is  clear and  well studied  that the RC  stars follow  this barred
Galactic feature, but in the RR  Lyrae case there is no clear evidence
for the  same trend.  On one hand  \cite{ogleIIIRRL} with  OGLE-III RR
Lyrae stars claim the existence  of the barred structure rotated about
$30^\circ$ with respect  to the line of sight between  the Sun and the
Galactic center.   On the  other hand \cite{dekany13}  completely rule
out this possibility using the same dataset, but (crucially) including
the near-IR results of the VVV Survey.

\begin{figure*}%[h!]
\centering
\includegraphics[scale=0.45]{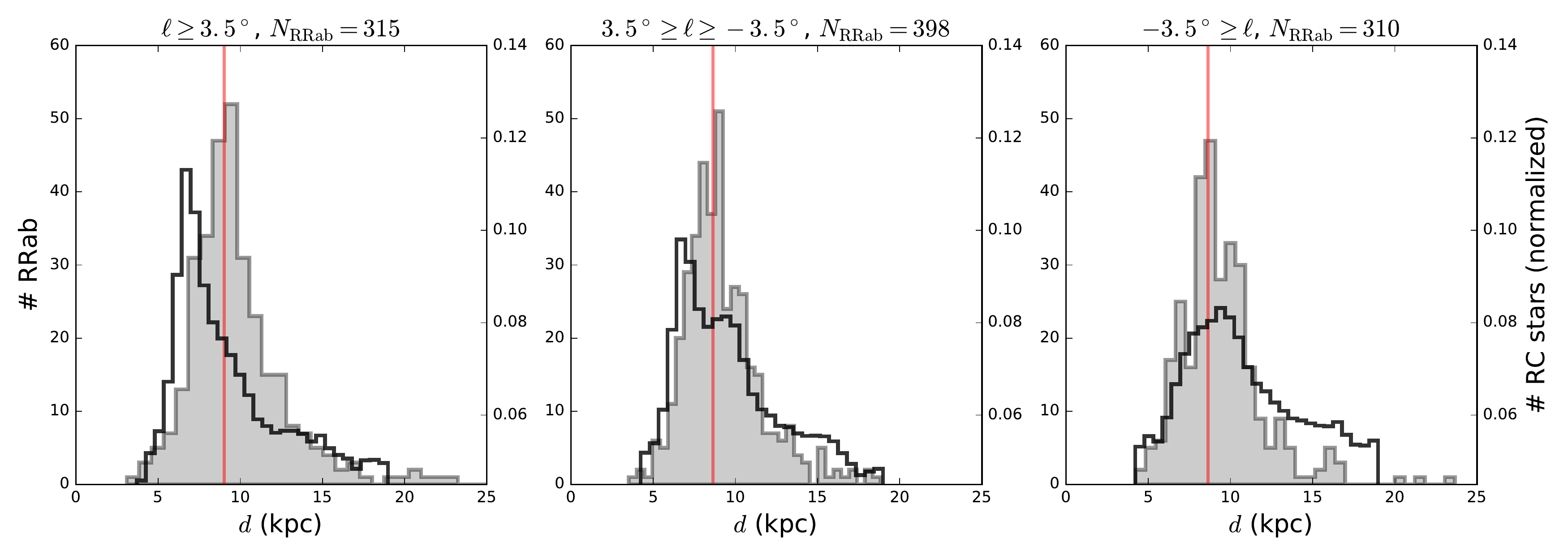}
\caption{Histogram  of distances  of RR  Lyrae (filled)  and  RC stars
  (steps) as function of Galactic latitude ($\ell$). The distributions
  of the RC stars include also  the underlying RGB, but since those do
  not affect the position of the RC the distributions are suitable for
  our comparison  purposes. The total number  of RC stars  in the same
  areas overwhelms the number of  RR Lyrae, thus the histogram showing
  their   distribution   in  distance   was   normalized  for   better
  visualization.  The vertical  line  represents the  RR Lyrae  median
  distance of each region.}
\label{fig:xshape}
\end{figure*}

We have used our catalogue to  compare the distribution of RR Lyrae at
low Galactic  latitude with the distribution  of RC stars  in the same
analyzed tiles. Both catalogues  were divided in three longitude bins:
$-10^\circ < \ell  < -3.5^\circ$; $-3.5^\circ < \ell  < 3.5^\circ$ and
$3.5^\circ <  \ell <  10^\circ$. The RC  stars were selected  with the
same technique  described in \cite{minniti11}  with magnitudes $K_{\rm
  s} <  15$, effectively limiting our  study to RC  stars at distances
closer than $ \sim 20$ kpc.  The distributions of the RC stars include
also the contribution  of the underlying RGB. The  RGB does not change
the position  of the RC, thus  the distributions are  suitable for our
comparison  purposes.   Assuming an  intrinsic  RC absolute  magnitude
$M_{K_{\rm s}} =  -1.55$ and an intrinsic RC  color $(J-K_{\rm s})_0 =
0.68$, as given by \cite{gonzalez11}  for Baade's window RC stars, the
distance equation yields:

\begin{equation}
  \mu = - 5 + 5 \log{d \text{ (pc)}}= K_{\rm s} - 0.73 (J - K_{\rm s}) + 2.05,
\end{equation}
where the \cite{cardelli89} extinction law was assumed.

Fig.~\ref{fig:xshape} shows  the result of the  comparison between the
distance distribution of RC and  RR Lyrae stars. The red vertical line
shows the mean distance value of a single Gaussian fit to the RR Lyrae
in each  longitude bin, namely  $d_{\rm{RRL}} \sim 9.01$,  $8.63$, and
$8.98$ kpc, from positive  to negative longitudes, respectively.  With
associated standard deviation of $\sigma_{\rm{RRL}} \sim 1.36$, $1.31$
and  $1.35$,  respectively.   Clearly,  the variation  in  the  median
distance across the longitude direction is negligible for the RR Lyrae
distribution.   RC stars,  on  the  contrary, show  a  single peak  at
$d_{\rm{RC}}  \sim  6.8$ kpc  at  positive  longitudes,  two peaks  at
$d_{\rm{RC}} \sim  6.8$ and  $9.5$, kpc across  the minor axis,  and a
single peak at $d_{\rm{RC}} \sim  9.4$ kpc at negative longitudes.  In
all three cases, a two-sample Kolmogorov-Smirnov test reveals that the
distributions  of RC  and RR  Lyrae stars  are indeed  different, with
higher than  98.9\% probability.  This  strongly suggests that  the RC
stars (but not the RR Lyrae)  follow the main Galactic bar, flaring up
into a peanut (X-shape) far  away from the Galactic plane.  The marked
difference in the  distance distribution of RR Lyrae  variables and RC
stars confirms,  at low latitudes, the  conclusion by \cite{dekany13},
that  RR Lyrae  and RC  stars trace  two different  components  in the
bulge.

\section{Summary}
\label{sec:summary}

A search  for RR Lyrae stars was  performed in more than  $\sim 47$ sq
deg  in the  outer parts  of the  Galactic bulge  observed by  the VVV
Survey.  In total, more than 1000 fundamental mode RR Lyrae stars were
found in this area, with  a completeness level estimated in $80\%$ for
$K_{\rm s}\lesssim15$~mag. We have analyzed their periods, amplitudes,
light curve  shapes, and 3D  positions within the Galaxy.  This sample
allow us to  compare the distribution along the  Galactic longitude of
RR Lyrae  and RC stars, resulting in  (statistically very significant)
differences  of  more  than   $1.5$~kpc  between  the  peaks  of  both
distributions. These differences  prevail along the Galactic latitudes
observed by the  VVV Survey that shows an  unchanged RR Lyrae distance
distribution and a  moving RC distribution tracing the  Milky Way bar.
This   result  fully   supports  the   work  of   \cite{dekany13}  and
\cite{kunder16} which  postulates a spheroidal distribution  of the RR
Lyrae stars in  the Galactic bulge, not tracing the  strong bar of the
RC  stars.  A  complete view  of the  RR Lyrae  stars over  the entire
Galactic bulge will  be unveiled when fully automatic  searches in the
VVV Survey area is completed \citep{catelan13b, angeloni14}.

\begin{acknowledgements}
We gratefully acknowledge  the use of data from  the ESO Public Survey
program ID  179.B-2002 taken with  the VISTA telescope,  data products
from the  Cambridge Astronomical Survey Unit. Support  for the authors
is provided by  the BASAL CATA Center for  Astrophysics and Associated
Technologies through  grant PFB-06, and the Ministry  for the Economy,
Development,  and Tourism's  Programa Iniciativa  Cient\'ifica Milenio
through  grant  IC120009,  awarded  to  the  Millennium  Institute  of
Astrophysics (MAS).   D.M. and M.Z. acknowledge  support from FONDECYT
Regular grants No. 1130196 and 1150345, respectively. Support for this
project is provided by CONICYT's PCI program through grant DPI20140066
F.G., C.N., and M.C.   acknowledge support from FONDECYT regular grant
No. 1141141.  C.N.  and  F.  G.  acknowledge support from CONICYT-PCHA
Doctorado  and Mag\'ister  Nacional  2015-21151643 and  2014-22141509,
respectively.  R.K.S.   acknowledges support from  CNPq/Brazil through
projects 310636/2013-2  and 481468/2013-7.  We  gratefully acknowledge
the use  of IPython, Astropy, AstroML, Matplotlib,  TOPCAT, and ALADIN
sky atlas.

\end{acknowledgements}

\begin{appendix}
\section{List of VVV RRab variables}
\label{app:data}

Table A.1 lists the main parameters of the 1019 ab-type RR Lyrae stars
discovered in  this work.   For each object  we provide the  VVV name,
Equatorial   and   Galactic   coordinates,   mean   $K_{\rm   s}$-band
weighted-magnitude,  period, amplitude,  and heliocentric  distance In
Table A.2 we  list the VVV RR Lyrae matching  variables in the General
Catalogue of Variable Stars (GCVS).

\clearpage

\begin{table*}[ht!]
\centering
\caption[]{List of VVV  RRab variables.  VVV IDs marked  with a single
  asterisk are the objects matching a variable in the General Cataloge
  of Variable Stars (GCVS). The corresponding GCVS names are presented
  in Table A.2. Three RRab  stars previously discovered by OGLE-IV are
  marked  with  double   asterisks,  namely:  J181632.17$-$334319.8  =
  OGLE-BLG-RRLYR-35577,  J181727.29$-$335532.1  = OGLE-BLG-RRLYR-35765
  and J181745.11$-$333025.2 = OGLE-BLG-RRLYR-35810.
\label{tab:tiles}}
\begin{tabular}{lcccccccr}
\hline \hline
VVV ID & RA (J2000.0)  & DEC (J2000.0) & $\ell$ & $b$ &  $\langle K_{\rm s} \rangle$ & $P$ & Amplitude & $d$~~~ \\
     & hh:mm:ss.ss    & dd:mm:ss.s   &  (deg) & (deg) &   (mag)       & (days) & (mag) & (kpc)   \\
\hline
J175459.11$-$413137.1 & 17:54:59.11 & $-$41:31:37.1 & $-$9.908 & $-$8.039 & 14.641 & 0.472833 & 0.30 & 10.0 \\
J175601.56$-$412549.0 & 17:56:01.56 & $-$41:25:49.0 & $-$9.727 & $-$8.161 & 13.998 & 0.656728 & 0.26 & 8.8 \\
J175609.49$-$410752.7 & 17:56:09.49 & $-$41:07:52.7 & $-$9.451 & $-$8.037 & 14.009 & 0.541568 & 0.37 & 8.0 \\
J175628.40$-$411745.8 & 17:56:28.40 & $-$41:17:45.8 & $-$9.567 & $-$8.169 & 13.651 & 0.517874 & 0.32 & 6.5 \\
J175718.55$-$412623.3 & 17:57:18.55 & $-$41:26:23.3 & $-$9.617 & $-$8.376 & 14.262 & 0.481343 & 0.35 & 8.4 \\
J175727.54$-$414338.0 & 17:57:27.54 & $-$41:43:38.0 & $-$9.858 & $-$8.540 & 14.875 & 0.589605 & 0.34 & 12.7 \\
\hline
\end{tabular}
\end{table*}

\addtocounter{table}{-1}

\begin{table*}
\centering
\caption[]{{\it continued}}
\begin{tabular}{lcccccccr}
\hline \hline
VVV ID & RA (J2000.0)  & DEC (J2000.0) & $\ell$ & $b$ &  $\langle K_{\rm s} \rangle$ & $P$ & Amplitude & $d$~~~ \\
     & hh:mm:ss.ss    & dd:mm:ss.s   &  (deg) & (deg) &   (mag)       & (days) & (mag) & (kpc)   \\
\hline
J180334.27$-$421303.7 & 18:03:34.27 & $-$42:13:03.7 & $-$9.746 & $-$9.772 & 15.045 & 0.523470 & 0.34 & 12.9 \\
J180335.31$-$395348.3 & 18:03:35.31 & $-$39:53:48.3 & $-$7.671 & $-$8.682 & 13.765 & 0.539700 & 0.29 &  7.1 \\
J180342.46$-$392615.5\,* & 18:03:42.46 & $-$39:26:15.5 & $-$7.251 & $-$8.484 & 14.093 & 0.390447 & 0.41 & 6.9 \\
J180345.50$-$420853.7 & 18:03:45.50 & $-$42:08:53.7 & $-$9.668 & $-$9.770 & 14.076 & 0.511941 & 0.33 & 8.0 \\
J180346.93$-$390232.0 & 18:03:46.93 & $-$39:02:32.0 & $-$6.892 & $-$8.310 & 14.304 & 0.508466 & 0.31 & 8.9 \\
J180351.95$-$390948.3\,* & 18:03:51.95 & $-$39:09:48.3 & $-$6.992 & $-$8.381 & 13.844 & 0.534605 & 0.30 & 7.3 \\
\hline
\end{tabular}
\end{table*}

\addtocounter{table}{-1}

\begin{table*}
\centering
\caption[]{{\it continued}}
\begin{tabular}{lcccccccr}
\hline \hline
VVV ID & RA (J2000.0)  & DEC (J2000.0) & $\ell$ & $b$ &  $\langle K_{\rm s} \rangle$ & $P$ & Amplitude & $d$~~~ \\
     & hh:mm:ss.ss    & dd:mm:ss.s   &  (deg) & (deg) &   (mag)       & (days) & (mag) & (kpc)   \\
\hline
J180713.57$-$413121.7 & 18:07:13.57 & $-$41:31:21.7 & $-$8.801 & $-$10.052 & 14.443 & 0.519276 & 0.21 & 9.6 \\
J180715.53$-$382158.1 & 18:07:15.53 & $-$38:21:58.1 & $-$5.966 & $-$8.589 & 14.914 & 0.445800 & 0.23 & 11.1 \\
J180718.20$-$400044.4 & 18:07:18.20 & $-$40:00:44.4 & $-$7.436 & $-$9.365 & 14.176 & 0.729320 & 0.18 & 10.1 \\
J180720.63$-$410313.3 & 18:07:20.63 & $-$41:03:13.3 & $-$8.368 & $-$9.854 & 13.387 & 0.816446 & 0.17 & 7.3 \\
J180721.68$-$401440.9 & 18:07:21.68 & $-$40:14:40.9 & $-$7.639 & $-$9.483 & 14.370 & 0.491681 & 0.28 & 9.0 \\
J180728.98$-$374352.8 & 18:07:28.98 & $-$37:43:52.8 & $-$5.378 & $-$8.330 & 14.306 & 0.611396 & 0.22 & 9.8 \\
\hline
\end{tabular}
\end{table*}

\addtocounter{table}{-1}

\begin{table*}
\centering
\caption[]{{\it continued}}
\begin{tabular}{lcccccccr}
\hline \hline
VVV ID & RA (J2000.0)  & DEC (J2000.0) & $\ell$ & $b$ &  $\langle K_{\rm s} \rangle$ & $P$ & Amplitude & $d$~~~ \\
     & hh:mm:ss.ss    & dd:mm:ss.s   &  (deg) & (deg) &   (mag)       & (days) & (mag) & (kpc)   \\
\hline
J181018.92$-$372946.2 & 18:10:18.92 & $-$37:29:46.2 & $-$4.903 & $-$8.716 & 14.632 & 0.474201 & 0.25 & 10.0 \\
J181020.74$-$400650.8 & 18:10:20.74 & $-$40:06:50.8 & $-$7.255 & $-$9.928 & 14.434 & 0.531426 & 0.32 & 9.7 \\
J181022.56$-$404726.1 & 18:10:22.56 & $-$40:47:26.1 & $-$7.863 & $-$10.242 & 14.432 & 0.457733 & 0.30 & 8.9 \\
J181023.42$-$364335.2 & 18:10:23.42 & $-$36:43:35.2 & $-$4.207 & $-$8.372 & 14.146 & 0.652835 & 0.25 & 9.4 \\
J181025.95$-$375450.8 & 18:10:25.95 & $-$37:54:50.8 & $-$5.267 & $-$8.93 & 14.962 & 0.474816 & 0.36 & 11.7 \\
J181027.91$-$362807.1 & 18:10:27.91 & $-$36:28:07.1 & $-$3.969 & $-$8.265 & 13.054 & 0.814930 & 0.29 & 6.2 \\
\hline
\end{tabular}
\end{table*}

\addtocounter{table}{-1}

\begin{table*}
\centering
\caption[]{{\it continued}}
\begin{tabular}{lcccccccr}
\hline \hline
VVV ID & RA (J2000.0)  & DEC (J2000.0) & $\ell$ & $b$ &  $\langle K_{\rm s} \rangle$ & $P$ & Amplitude & $d$~~~ \\
     & hh:mm:ss.ss    & dd:mm:ss.s   &  (deg) & (deg) &   (mag)       & (days) & (mag) & (kpc)   \\
\hline
J181230.46$-$375218.9 & 18:12:30.46 & $-$37:52:18.9 & $-$5.038 & $-$9.274 & 14.674 & 0.603464 & 0.34 & 11.6 \\
J181230.83$-$395323.9 & 18:12:30.83 & $-$39:53:23.9 & $-$6.860 & $-$10.196 & 13.954 & 0.652405 & 0.33 & 8.6 \\
J181237.48$-$350757.4\,* & 18:12:37.48 & $-$35:07:57.4 & $-$2.569 & $-$8.032 & 13.692 & 0.477544 & 0.35 & 6.4 \\
J181240.59$-$364330.2\,* & 18:12:40.59 & $-$36:43:30.2 & $-$3.991 & $-$8.777 & 14.586 & 0.445500 & 0.28 & 9.5 \\
J181241.48$-$370833.3 & 18:12:41.48 & $-$37:08:33.3 & $-$4.365 & $-$8.972 & 14.702 & 0.571893 & 0.27 & 11.5 \\
J181245.83$-$375541.8 & 18:12:45.83 & $-$37:55:41.8 & $-$5.066 & $-$9.345 & 14.875 & 0.606211 & 0.25 & 12.9 \\
\hline
\end{tabular}
\end{table*}

\addtocounter{table}{-1}

\begin{table*}
\centering
\caption[]{{\it continued}}
\begin{tabular}{lcccccccr}
\hline \hline
VVV ID & RA (J2000.0)  & DEC (J2000.0) & $\ell$ & $b$ &  $\langle K_{\rm s} \rangle$ & $P$ & Amplitude & $d$~~~ \\
     & hh:mm:ss.ss    & dd:mm:ss.s   &  (deg) & (deg) &   (mag)       & (days) & (mag) & (kpc)   \\
\hline
J181447.56$-$374749.2 & 18:14:47.56 & $-$37:47:49.2 & $-$4.762 & $-$9.642 & 13.129 & 0.561060 & 0.28 & 5.3 \\
J181458.41$-$351324.8\,* & 18:14:58.41 & $-$35:13:24.8 & $-$2.427 & $-$8.500 & 14.389 & 0.520159 & 0.30 & 9.4 \\
J181505.11$-$374522.4 & 18:15:05.11 & $-$37:45:22.4 & $-$4.699 & $-$9.675 & 14.087 & 0.781305 & 0.32 & 10.1 \\
J181505.80$-$390448.3 & 18:15:05.80 & $-$39:04:48.3 & $-$5.897 & $-$10.274 & 14.016 & 0.644749 & 0.32 & 8.8 \\
J181506.49$-$345411.4\,* & 18:15:06.49 & $-$34:54:11.4 & $-$2.126 & $-$8.377 & 15.189 & 0.536993 & 0.30 & 14.0 \\
J181507.09$-$382029.1 & 18:15:07.09 & $-$38:20:29.1 & $-$5.226 & $-$9.945 & 13.714 & 0.620582 & 0.28 & 7.4 \\
\hline
\end{tabular}
\end{table*}

\addtocounter{table}{-1}

\begin{table*}
\centering
\caption[]{{\it continued}}
\begin{tabular}{lcccccccr}
\hline \hline
VVV ID & RA (J2000.0)  & DEC (J2000.0) & $\ell$ & $b$ &  $\langle K_{\rm s} \rangle$ & $P$ & Amplitude & $d$~~~ \\
     & hh:mm:ss.ss    & dd:mm:ss.s   &  (deg) & (deg) &   (mag)       & (days) & (mag) & (kpc)   \\
\hline
J181713.34$-$372529.4 & 18:17:13.34 & $-$37:25:29.4 & $-$4.205 & $-$9.903 & 14.074 & 0.611086 & 0.16 & 8.8 \\
J181718.08$-$375812.6 & 18:17:18.08 & $-$37:58:12.6 & $-$4.693 & $-$10.162 & 14.277 & 0.577264 & 0.33 & 9.4 \\
J181723.79$-$371424.2 & 18:17:23.79 & $-$37:14:24.2 & $-$4.022 & $-$9.851 & 14.357 & 0.520143 & 0.32 & 9.2 \\
J181724.11$-$371520.9 & 18:17:24.11 & $-$37:15:20.9 & $-$4.035 & $-$9.859 & 14.776 & 0.556967 & 0.29 & 11.7 \\
J181724.66$-$345352.6 & 18:17:24.66 & $-$34:53:52.6 & $-$1.903 & $-$8.795 & 14.332 & 0.462888 & 0.29 & 8.5 \\
J181725.42$-$334742.4\,* & 18:17:25.42 & $-$33:47:42.4 & $-$0.909 & $-$8.295 & 13.950 & 0.534068 & 0.20 & 7.7 \\
\hline
\end{tabular}
\end{table*}

\addtocounter{table}{-1}

\begin{table*}
\centering
\caption[]{{\it continued}}
\begin{tabular}{lcccccccr}
\hline \hline
VVV ID & RA (J2000.0)  & DEC (J2000.0) & $\ell$ & $b$ &  $\langle K_{\rm s} \rangle$ & $P$ & Amplitude & $d$~~~ \\
     & hh:mm:ss.ss    & dd:mm:ss.s   &  (deg) & (deg) &   (mag)       & (days) & (mag) & (kpc)   \\
\hline
J181923.70$-$352208.4\,* & 18:19:23.70 & $-$35:22:08.4 & $-$2.143 & $-$9.370 & 15.270 & 0.406951 & 0.34 & 12.5 \\
J181926.21$-$362655.4 & 18:19:26.21 & $-$36:26:55.4 & $-$3.118 & $-$9.861 & 14.527 & 0.544556 & 0.23 & 10.2 \\
J181926.95$-$345327.4\,* & 18:19:26.95 & $-$34:53:27.4 & $-$1.705 & $-$9.164 & 13.466 & 0.601815 & 0.27 & 6.5 \\
J181934.58$-$371913.7 & 18:19:34.58 & $-$37:19:13.7 & $-$3.898 & $-$10.275 & 13.895 & 0.645403 & 0.24 & 8.3 \\
J181934.72$-$355159.1 & 18:19:34.72 & $-$35:51:59.1 & $-$2.577 & $-$9.626 & 13.798 & 0.589247 & 0.25 & 7.5 \\
J181938.68$-$323528.0 & 18:19:38.68 & $-$32:35:28.0 &  0.388 & $-$8.159 & 14.433 & 0.829071 & 0.21 & 12.3 \\
\hline
\end{tabular}
\end{table*}

\addtocounter{table}{-1}

\begin{table*}
\centering
\caption[]{{\it continued}}
\begin{tabular}{lcccccccr}
\hline \hline
VVV ID & RA (J2000.0)  & DEC (J2000.0) & $\ell$ & $b$ &  $\langle K_{\rm s} \rangle$ & $P$ & Amplitude & $d$~~~ \\
     & hh:mm:ss.ss    & dd:mm:ss.s   &  (deg) & (deg) &   (mag)       & (days) & (mag) & (kpc)   \\
\hline
J182154.23$-$345156.0 & 18:21:54.23 & $-$34:51:56.0 & $-$1.454 & $-$9.603 & 14.256 & 0.542029 & 0.24 & 9.0 \\
J182157.52$-$322432.2 & 18:21:57.52 & $-$32:24:32.2 &  0.775 & $-$8.512 & 14.413 & 0.570558 & 0.26 & 9.9 \\
J182200.35$-$322139.1\,* & 18:22:00.35 & $-$32:21:39.1 &  0.823 & $-$8.499 & 15.004 & 0.552370 & 0.35 & 13.0 \\
J182201.27$-$334837.7 & 18:22:01.27 & $-$33:48:37.7 & $-$0.486 & $-$9.153 & 15.140 & 0.569133 & 0.21 & 14.1 \\
J182204.24$-$350828.6 & 18:22:04.24 & $-$35:08:28.6 & $-$1.689 & $-$9.756 & 13.981 & 0.560444 & 0.25 & 8.0 \\
J182209.13$-$340440.9\,* & 18:22:09.13 & $-$34:04:40.9 & $-$0.716 & $-$9.297 & 14.331 & 0.540690 & 0.30 & 9.3 \\
\hline
\end{tabular}
\end{table*}

\addtocounter{table}{-1}

\begin{table*}
\centering
\caption[]{{\it continued}}
\begin{tabular}{lcccccccr}
\hline \hline
VVV ID & RA (J2000.0)  & DEC (J2000.0) & $\ell$ & $b$ &  $\langle K_{\rm s} \rangle$ & $P$ & Amplitude & $d$~~~ \\
     & hh:mm:ss.ss    & dd:mm:ss.s   &  (deg) & (deg) &   (mag)       & (days) & (mag) & (kpc)   \\
\hline
J182400.52$-$315930.3 & 18:24:00.52 & $-$31:59:30.3 &  1.349 & $-$8.711 & 15.387 & 0.530904 & 0.33 & 15.3 \\
J182402.88$-$330313.4 & 18:24:02.88 & $-$33:03:13.4 &  0.391 & $-$9.193 & 14.490 & 0.603573 & 0.30 & 10.6 \\
J182405.17$-$342148.9 & 18:24:05.17 & $-$34:21:48.9 & $-$0.795 & $-$9.782 & 13.999 & 0.537014 & 0.26 & 7.9 \\
J182405.27$-$313547.9\,* & 18:24:05.27 & $-$31:35:47.9 &  1.714 & $-$8.549 & 14.168 & 0.515063 & 0.31 & 8.4 \\
J182410.61$-$325402.0 & 18:24:10.61 & $-$32:54:02.0 &  0.542 & $-$9.149 & 14.440 & 0.573694 & 0.26 & 10.1 \\
J182411.73$-$302708.1 & 18:24:11.73 & $-$30:27:08.1 &  2.757 & $-$8.055 & 14.620 & 0.473788 & 0.36 & 9.9 \\
\hline
\end{tabular}
\end{table*}

\addtocounter{table}{-1}

\begin{table*}
\centering
\caption[]{{\it continued}}
\begin{tabular}{lcccccccr}
\hline \hline
VVV ID & RA (J2000.0)  & DEC (J2000.0) & $\ell$ & $b$ &  $\langle K_{\rm s} \rangle$ & $P$ & Amplitude & $d$~~~ \\
     & hh:mm:ss.ss    & dd:mm:ss.s   &  (deg) & (deg) &   (mag)       & (days) & (mag) & (kpc)   \\
\hline
J182607.09$-$321309.4 & 18:26:07.09 & $-$32:13:09.4 &  1.345 & $-$9.213 & 14.328 & 0.490855 & 0.30 & 8.8 \\
J182607.78$-$312211.2\,* & 18:26:07.78 & $-$31:22:11.2 &  2.116 & $-$8.837 & 13.731 & 0.595912 & 0.27 & 7.3 \\
J182608.84$-$303346.1 & 18:26:08.84 & $-$30:33:46.1 &  2.848 & $-$8.480 & 13.765 & 0.493715 & 0.24 & 6.7 \\
J182609.06$-$334753.2 & 18:26:09.06 & $-$33:47:53.2 & $-$0.088 & $-$9.916 & 14.122 & 0.697725 & 0.31 & 9.6 \\
J182609.53$-$321657.6 & 18:26:09.53 & $-$32:16:57.6 &  1.291 & $-$9.248 & 14.581 & 0.534625 & 0.35 & 10.4 \\
J182609.71$-$302815.0 & 18:26:09.71 & $-$30:28:15.0 &  2.933 & $-$8.442 & 13.827 & 0.643770 & 0.37 & 8.0 \\
\hline
\end{tabular}
\end{table*}

\addtocounter{table}{-1}

\begin{table*}
\centering
\caption[]{{\it continued}}
\begin{tabular}{lcccccccr}
\hline \hline
VVV ID & RA (J2000.0)  & DEC (J2000.0) & $\ell$ & $b$ &  $\langle K_{\rm s} \rangle$ & $P$ & Amplitude & $d$~~~ \\
     & hh:mm:ss.ss    & dd:mm:ss.s   &  (deg) & (deg) &   (mag)       & (days) & (mag) & (kpc)   \\
\hline
J182828.73$-$330519.7\,* & 18:28:28.73 & $-$33:05:19.7 &  0.775 & $-$10.042 & 13.838 & 0.580134 & 0.32 & 7.6 \\
J182829.11$-$295836.2 & 18:28:29.11 & $-$29:58:36.2 &  3.606 & $-$8.670 & 14.601 & 0.544609 & 0.32 & 10.6 \\
J182829.60$-$285430.9 & 18:28:29.60 & $-$28:54:30.9 &  4.574 & $-$8.196 & 14.077 & 0.475703 & 0.24 & 7.7 \\
J182829.65$-$303455.0 & 18:28:29.65 & $-$30:34:55.0 &  3.059 & $-$8.941 & 14.385 & 0.635021 & 0.26 & 10.4 \\
J182831.16$-$283358.6 & 18:28:31.16 & $-$28:33:58.6 &  4.886 & $-$8.048 & 14.523 & 0.501339 & 0.35 & 9.8 \\
J182831.99$-$311120.0\,* & 18:28:31.99 & $-$31:11:20.0 &  2.511 & $-$9.217 & 15.116 & 0.506837 & 0.26 & 13.1 \\
\hline
\end{tabular}
\end{table*}

\addtocounter{table}{-1}

\begin{table*}
\centering
\caption[]{{\it continued}}
\begin{tabular}{lcccccccr}
\hline \hline
VVV ID & RA (J2000.0)  & DEC (J2000.0) & $\ell$ & $b$ &  $\langle K_{\rm s} \rangle$ & $P$ & Amplitude & $d$~~~ \\
     & hh:mm:ss.ss    & dd:mm:ss.s   &  (deg) & (deg) &   (mag)       & (days) & (mag) & (kpc)   \\
\hline
J183045.33$-$322237.8\,* & 18:30:45.33 & $-$32:22:37.8 &  1.638 & $-$10.162 & 14.691 & 0.490333 & 0.34 & 10.5 \\
J183048.73$-$310954.4\,* & 18:30:48.73 & $-$31:09:54.4 &  2.750 & $-$9.644 & 14.274 & 0.491565 & 0.35 & 8.6 \\
J183051.28$-$320058.4 & 18:30:51.28 & $-$32:00:58.4 &  1.977 & $-$10.024 & 14.302 & 0.549064 & 0.28 & 9.2 \\
J183054.75$-$292227.6 & 18:30:54.75 & $-$29:22:27.6 &  4.389 & $-$8.876 & 14.108 & 0.578777 & 0.30 & 8.6 \\
J183055.00$-$285232.1 & 18:30:55.00 & $-$28:52:32.1 &  4.842 & $-$8.656 & 14.646 & 0.698880 & 0.28 & 12.4 \\
J183056.63$-$323329.3 & 18:30:56.63 & $-$32:33:29.3 &  1.490 & $-$10.277 & 14.543 & 0.624707 & 0.24 & 11.1 \\
\hline
\end{tabular}
\end{table*}

\addtocounter{table}{-1}

\begin{table*}
\centering
\caption[]{{\it continued}}
\begin{tabular}{lcccccccr}
\hline \hline
VVV ID & RA (J2000.0)  & DEC (J2000.0) & $\ell$ & $b$ &  $\langle K_{\rm s} \rangle$ & $P$ & Amplitude & $d$~~~ \\
     & hh:mm:ss.ss    & dd:mm:ss.s   &  (deg) & (deg) &   (mag)       & (days) & (mag) & (kpc)   \\
\hline
J183307.09$-$305707.5 & 18:33:07.09 & $-$30:57:07.5 &  3.163 & $-$9.996 & 14.411 & 0.521722 & 0.24 & 9.5 \\
J183308.14$-$294313.4 & 18:33:08.14 & $-$29:43:13.4 &  4.290 & $-$9.463 & 14.784 & 0.489764 & 0.32 & 11.0 \\
J183310.92$-$285537.8 & 18:33:10.92 & $-$28:55:37.8 &  5.017 & $-$9.124 & 14.227 & 0.643360 & 0.28 & 9.7 \\
J183317.23$-$264230.6 & 18:33:17.23 & $-$26:42:30.6 &  7.041 & $-$8.166 & 14.440 & 0.549345 & 0.29 & 9.9 \\
J183317.54$-$283543.8 & 18:33:17.54 & $-$28:35:43.8 &  5.329 & $-$9.000 & 14.733 & 0.574082 & 0.26 & 11.6 \\
J183320.37$-$293233.8 & 18:33:20.37 & $-$29:32:33.8 &  4.471 & $-$9.425 & 14.814 & 0.477637 & 0.40 & 11.0 \\
\hline
\end{tabular}
\end{table*}

\addtocounter{table}{-1}

\begin{table*}
\centering
\caption[]{{\it continued}}
\begin{tabular}{lcccccccr}
\hline \hline
VVV ID & RA (J2000.0)  & DEC (J2000.0) & $\ell$ & $b$ &  $\langle K_{\rm s} \rangle$ & $P$ & Amplitude & $d$~~~ \\
     & hh:mm:ss.ss    & dd:mm:ss.s   &  (deg) & (deg) &   (mag)       & (days) & (mag) & (kpc)   \\
\hline
J183532.33$-$264853.6 & 18:35:32.33 & $-$26:48:53.6 &  7.169 & $-$8.664 & 14.838 & 0.490952 & 0.42 & 11.3 \\
J183536.59$-$271717.0 & 18:35:36.59 & $-$27:17:17.0 &  6.746 & $-$8.886 & 13.882 & 0.494130 & 0.33 & 7.1 \\
J183538.81$-$270026.8 & 18:35:38.81 & $-$27:00:26.8 &  7.005 & $-$8.770 & 14.268 & 0.501819 & 0.33 & 8.6 \\
J183540.25$-$282947.5 & 18:35:40.25 & $-$28:29:47.5 &  5.651 & $-$9.427 & 15.527 & 0.524981 & 0.37 & 16.3 \\
J183549.58$-$265310.4 & 18:35:49.58 & $-$26:53:10.4 &  7.133 & $-$8.753 & 13.977 & 0.531082 & 0.29 & 7.7 \\
J183551.71$-$280526.9 & 18:35:51.71 & $-$28:05:26.9 &  6.039 & $-$9.288 & 14.130 & 0.534291 & 0.32 & 8.4 \\
\hline
\end{tabular}
\end{table*}

\addtocounter{table}{-1}

\begin{table*}
\centering
\caption[]{{\it continued}}
\begin{tabular}{lcccccccr}
\hline \hline
VVV ID & RA (J2000.0)  & DEC (J2000.0) & $\ell$ & $b$ &  $\langle K_{\rm s} \rangle$ & $P$ & Amplitude & $d$~~~ \\
     & hh:mm:ss.ss    & dd:mm:ss.s   &  (deg) & (deg) &   (mag)       & (days) & (mag) & (kpc)   \\
\hline
J183804.87$-$263637.9 & 18:38:04.87 & $-$26:36:37.9 &  7.607 & $-$9.085 & 13.827 & 0.558887 & 0.27 & 7.4 \\
J183805.70$-$285735.7 & 18:38:05.70 & $-$28:57:35.7 &  5.461 & $-$10.106 & 14.879 & 0.451048 & 0.39 & 11.0 \\
J183806.30$-$275403.1 & 18:38:06.30 & $-$27:54:03.1 &  6.431 & $-$9.651 & 12.921 & 0.619721 & 0.25 & 5.1 \\
J183807.96$-$251732.7 & 18:38:07.96 & $-$25:17:32.7 &  8.811 & $-$8.518 & 13.542 & 0.487524 & 0.32 & 6.0 \\
J183808.34$-$255105.2 & 18:38:08.34 & $-$25:51:05.2 &  8.303 & $-$8.765 & 14.117 & 0.523390 & 0.25 & 8.2 \\
J183808.50$-$251840.5 & 18:38:08.50 & $-$25:18:40.5 &  8.795 & $-$8.528 & 14.690 & 0.463188 & 0.33 & 10.2 \\
\hline
\end{tabular}
\end{table*}

\addtocounter{table}{-1}

\begin{table*}
\centering
\caption[]{{\it continued}}
\begin{tabular}{lcccccccr}
\hline \hline
VVV ID & RA (J2000.0)  & DEC (J2000.0) & $\ell$ & $b$ &  $\langle K_{\rm s} \rangle$ & $P$ & Amplitude & $d$~~~ \\
     & hh:mm:ss.ss    & dd:mm:ss.s   &  (deg) & (deg) &   (mag)       & (days) & (mag) & (kpc)   \\
\hline
J184102.15$-$265412.7 & 18:41:02.15 & $-$26:54:12.7 &  7.629 & $-$9.806 & 14.674 & 0.513695 & 0.31 & 10.7 \\
J184125.74$-$261120.4 & 18:41:25.74 & $-$26:11:20.4 &  8.321 & $-$9.577 & 14.278 & 0.568870 & 0.27 & 9.3 \\
J184138.49$-$245701.6 & 18:41:38.49 & $-$24:57:01.6 &  9.473 & $-$9.083 & 14.317 & 0.572793 & 0.26 & 9.5 \\
J184142.59$-$244830.3 & 18:41:42.59 & $-$24:48:30.3 &  9.610 & $-$9.035 & 14.385 & 0.576399 & 0.26 & 9.9 \\
J184147.89$-$254048.9 & 18:41:47.89 & $-$25:40:48.9 &  8.823 & $-$9.432 & 13.910 & 0.691290 & 0.27 & 8.6 \\
J184149.39$-$251819.0 & 18:41:49.39 & $-$25:18:19.0 &  9.168 & $-$9.274 & 13.812 & 0.642426 & 0.25 & 7.9 \\
\hline                                                                                                       
\end{tabular}                                                                                                
\end{table*}

\clearpage

\begin{table*}[ht!]
\centering
\caption[]{List of the 207 VVV RRab matching a variable in the General
  Cataloge of Variable  Stars (GCVS). VVV IDs and  the respective GCVS
  names are presented.}
\label{tab:gcvs}
\begin{tabular}{llllll}
\hline \hline
|~~VVV ID & GCVS name & |~~VVV ID & GCVS name & |~~VVV ID & GCVS name \\
\hline
|~~J2701020.22$-$411613.9 & V0462 CrA & |~~J2741313.37$-$375727.9 & IP    CrA & |~~J2755443.71$-$341148.1 & V2539 Sgr \\ 
|~~J2701149.18$-$420358.3 & V0463 CrA & |~~J2741423.44$-$381809.1 & V0536 CrA & |~~J2755521.67$-$312549.7 & V3212 Sgr \\
|~~J2702437.84$-$395238.8 & LM    CrA & |~~J2741507.38$-$341727.9 & V2874 Sgr & |~~J2755625.55$-$315420.3 & V3216 Sgr \\ 
|~~J2703434.23$-$413322.1 & V0469 CrA & |~~J2741640.08$-$341613.8 & V2878 Sgr & |~~J2755732.00$-$335639.4 & V3213 Sgr \\ 
|~~J2704032.86$-$393612.4 & LT    CrA & |~~J2742121.33$-$334742.4 & V2896 Sgr & |~~J2755903.99$-$320723.9 & V2540 Sgr \\ 
|~~J2705536.91$-$392615.5 & MV    CrA & |~~J2742130.69$-$361312.1 & V2889 Sgr & |~~J2760118.99$-$313547.9 & V3232 Sgr \\ 
|~~J2705759.29$-$390948.3 & MY    CrA & |~~J2742153.30$-$354403.3 & V2890 Sgr & |~~J2760406.80$-$325426.8 & V3242 Sgr \\
\hline
\end{tabular}
\end{table*}

\addtocounter{table}{-1}

\begin{table*}
\centering
\caption[]{{\it continued}}
\begin{tabular}{llllll}
\hline \hline
|~~VVV ID & GCVS name & |~~VVV ID & GCVS name & |~~VVV ID & GCVS name \\
\hline
|~~J2740455.20$-$353803.8 & V0702 Sgr & |~~J2754533.05$-$335723.9 & V3180 Sgr & |~~J2774809.03$-$314441.3 & V3490 Sgr \\ 
|~~J2740715.58$-$351629.9 & V0706 Sgr & |~~J2754558.25$-$320110.0 & V3188 Sgr & |~~J2775727.88$-$311535.4 & V2574 Sgr \\ 
|~~J2740722.74$-$352507.5 & V0707 Sgr & |~~J2754619.69$-$343303.0 & V3183 Sgr & |~~J2780805.50$-$310924.2 & V1625 Sgr \\ 
|~~J2740748.40$-$351726.8 & V0708 Sgr & |~~J2754708.67$-$341138.0 & V3186 Sgr & |~~J2780843.71$-$312018.5 & V2582 Sgr \\ 
|~~J2740804.45$-$334141.8 & V2844 Sgr & |~~J2754826.49$-$341617.6 & V1295 Sgr & |~~J2781001.76$-$313547.4 & V2583 Sgr \\ 
|~~J2741013.01$-$354614.0 & V2846 Sgr & |~~J2754847.92$-$314745.3 & V3193 Sgr & |~~J2781358.59$-$312612.9 & V3539 Sgr \\ 
|~~J2741048.75$-$340122.3 & V2852 Sgr & |~~J2754919.80$-$325428.2 & V3191 Sgr & |~~J2791305.39$-$291045.1 & V2597 Sgr \\ 
|~~J2741123.21$-$333359.5 & V2859 Sgr & |~~J2754928.83$-$311656.0 & V3197 Sgr & |~~J2792632.46$-$281530.6 & V1201 Sgr \\ 
|~~J2741224.96$-$352017.7 & V2857 Sgr & |~~J2755001.53$-$333533.8 & V3192 Sgr & |~~J2793223.91$-$274127.0 & V2374 Sgr \\ 
\hline       
\end{tabular}
\end{table*}

\end{appendix}

\end{document}